%% file: 14022003.tex

\documentclass[aip,onecolumn,amsmath,amssymb,reprint,fleqn,superbib,galley,twoside,superscriptaddress,a4paper,
unsortedaddress %%%%%% ??1??????????????????????¦Ë??????¨¢????????¦Ë?????????§³?
]{TAML-revtex4-1}

\usepackage{upgreek,pslatex,mathptm,mathptmx,setspace,pslatex}
\usepackage[english]{babel}
\allowdisplaybreaks

\renewcommand{\arraystretch}{1.2}

\def\mycmd{0} % ?????????Revised date??1-??Revised date; 0-???Revised date

\include{TAML-setting} %%%%%% ???????????? TAML ?????????
\begin{document}
\baselineskip=11.5pt plus.2pt minus.2pt
\parskip=0pt plus.2pt minus.2pt

%%%%%%%%%% ???????????3???????????????§µ????????????§µ??????????????????§Õ %%%%%%%%%%
\TitleOfPdf{\!\!\textrm{Space--time correlations in turbulent flow: A review}} % Input the Full title of paper ??2?? ??????????????
\ShortedTitle{Space--time correlations in turbulent flow: A review} %%% Short title for the page head ??3???????, ???????¨¹
\ShortedAuthor{J. M. Wallace} % W. Q. Liu, and W. Q. Liu,%%% ??4????????§Õ??????????????¨¹

%%%%%%%%%% ???????????? (For Editorial Office Only)?????¡À??? %%%%%%%%%%%%%%%%%%%%%%%%%%%%%%%%%%%
\NumberOfPaper{022003} %%% ??i?? ??????¦Ë????
\DoiOfPaper{10.1063/2.1402203} %%% ??ii??????DOI??10.1063/2.?????????, 13??????,????¦Ë???¦Ë?????§Ö???¦Ë?????????¦Ë??0??
\Received{18 February 2014} %%% ??iii??Recevied Date
\Revised{6 October 2013} % ??????????????????mycmd???????0??
\Accepted{24 February 2014} %%% ??iv??Accepted Date
%\PublishedOnline{10 November 2013} %%% ??v??Published Online Date
\VolNumberOfPaper{\textbf{4}, \the\NumberOfPaper ~(2014)} %%% ???????????? %%% ????????????
\include{TAML-title} %%%%%% ???????????? TAML ??????????????????????
%%%%%%%%%% ????????????? (For Editorial Office Only) ?????¡À??? %%%%%%%%%%%%%%%%%%%%%%%%%%%%%%%%%%%
\renewcommand{\figurename}{Fig.}
\renewcommand{\refname}{}

%%%%%% ??5?????????????????????????¦Ë???6??????????????????????AIP????APS????ÈÉ???COPY??????????????\mbox??
\author{\textrm{James M. Wallace}}%Weiqing\, Liu
\thanks{Corresponding author. Email: wallace@umd.edu.}
\affiliation{Burgers Program for Fluid Dynamics and Department of Mechanical Engineering,
University of Maryland, College Park, MD 20742, USA}%

\begin{abstract}
\setstretch{1.3}
\parbox{12.5cm}{\the\ManuscriptProcessing %%%%%% ???????
%%%%%%%%%%%% Begin of Abstract %%%%%%%%
\noindent \textbf{Abstract} %%%%%% ??7?? ?????
This paper reviews some of the principal uses, over almost seven decades, of correlations, in both Eulerian and Lagrangian frames of reference, of properties of turbulent flows at variable spatial locations and variable time instants. Commonly called space--time correlations, they have been fundamental to theories and models of turbulence as well as for the analyses of experimental and direct numerical simulation turbulence data.

%%%%%%%%%%%% End of Abstract %%%%
\the\Logo %%%%%% ???????

\noindent \textbf{Keywords} %%%%%% Keywords %%%%%% ??8?? ?????{}??
{turbulence, space-time correlation, Lagrangian and Eulerian correlations, direct interaction approximation, Taylor's hypothesis, elliptic approximation, particle dispersion}
} %%%%%%% ???????
\end{abstract}
%\keywords{Time {\textit{time \&}}}

\maketitle
\thispagestyle{newpaper}

%%%%%% ??9?? ?????? %%%%
{\setstretch{1.3}

\vspace*{-13mm}

\section{Introduction}

For many decades space--time correlations have been fundamental to
statistical theories of turbulence and modeling of some of its processes and to staple methods of data analysis for investigating turbulent flows. The Eulerian correlation coefficient of velocity components in stationary turbulent fields, fluctuating about their mean values, is defined most generally for two locations and two times as
\begin{eqnarray}\label{eq1}
R_{\rm E}({\pmb x}, {\pmb r}, \tau) = {\langle u_i ({\pmb x}, t_0) u_j({\pmb x} + {\pmb r}, t_0 + \tau) \rangle }/({ \sqrt{\langle {u_i^2({\pmb x}, t_0)}\rangle} \sqrt{\langle {u_j}^2({\pmb x} + {\pmb r}, t_0 + \tau)}\rangle}),
\end{eqnarray}
where the velocity fluctuations are denoted by $u_i$ and $u_j$ ($i,j = 1,2,3$), ${\pmb x}= (x_1, x_2, x_3)$ is a specified measurement location, ${\pmb x} + {\pmb r}=(x_1 + \Delta x_1, x_2 + \Delta x_2, x_3 + \Delta x_3)$ are locations with respect to ${\pmb x}$ that can be systematically varied, and $\tau$ is the time increment between the two times, $t_0$ and $t_0 + \tau$. Here, the numbered indices indicate the streamwise, wall normal and spanwise directions, respectively, and $\langle\cdot\rangle$ denotes the average of an ensemble of realizations. Correlation coefficients for other fluctuating turbulence properties, such as pressure, are expressed similarly.

Lagrangian correlation coefficients also can be defined for properties of fluid particles
that pass through Eulerian locations ${\pmb x}$ (in homogeneous planes of the flow) at times $t_0$ and travel along Lagrangian trajectories to arrive at positions ${\pmb x} + {\pmb r}(\tau)$ at times $t_0 + \tau$. In this case, the displacement vector, ${\pmb r}(t_0 + \tau)$, is a random variable describing the positions, at times $t_0 + \tau$, of the particles in the averaging ensemble with respect to the initial locations ${\pmb x}$ at times $t_0$ and that are different for each particle. Thus, for Lagrangian correlation coefficients, ${\pmb r}$ and $\tau$ are not varied independently, i.e., ${\pmb r}$ is a function of $\tau$. Such Lagrangian correlation coefficients are given by
\begin{eqnarray}\label{eq2}
{R_{\rm L}}_1({\pmb x}, \tau) = {\langle u_i ({\pmb x}, t_0)~u_j({\pmb x} + {\pmb r}(t_0 + \tau))\rangle}/({ \sqrt{\langle {u_i}^2({\pmb x}, t_0)\rangle} \sqrt{\langle {u_j}^2({\pmb x} + {\pmb r}(t_0 + \tau))}\rangle }).
\end{eqnarray}
Lagrangian correlations of two particles with some specified initial spatial separation can also be defined. In this case, the difference in the velocity components of the two particles are correlated for their pair of Lagrangian trajectories, and the correlation coefficient is given as
\begin{eqnarray}\label{eq3}
{R_{\rm L}}_2({\pmb x}, {\pmb r}(t_0), \tau) = { \langle d_i ({\pmb x}, {\pmb r}(t_0)) d_i({\pmb x} + {\pmb r}(t_0 + \tau)) \rangle}/({ \sqrt{\langle {d_i}^2({\pmb x}, {\pmb r}(t_0))}\rangle \sqrt{\langle {d_j}^2({\pmb x} + {\pmb r}(t_0 + \tau))}\rangle }),
\end{eqnarray}
where ${\pmb r}(t_0 + \tau)$ is the separation vector of the two particles at time $t_0 + \tau$, and $d_i = u_i ({\pmb x}) - u_i ({\pmb x} + {\pmb r}(t_0 + \tau))$.
Note that neither Eulerian nor Lagrangian correlation coefficients depend on the initial times $t_0$ for stationary turbulent fields. Multi-location, multi-time correlation coefficients have even been introduced\cite{1} with the availability of well resolved direct numerical simulations (DNS) of turbulence in space and time.

\section{Space--time correlations and turbulence theories and models}

Space--time correlations have played an important role in statistical theories of turbulence. The earliest of these is Taylor's\cite{2} celebrated treatment of the dispersion of fluid particles. He derived, for isotropic turbulence, an integral relationship between the single particle Lagrangian correlation coefficient and the mean square distance traveled, by an ensemble of fluid particles, from a specified location in the flow and in a particular coordinate direction.

Kraichnan's\cite{3,4,5} direct-interaction approximation (DIA) is formulated in terms of Eulerian space--time correlations, as is the related eddy damped quasi-normal Markovian (EDQNM) approximation. Kraichnan\cite{6} modeled the space--time correlations using his ``sweeping hypothesis'' that assumes that they are principally determined by a sweeping velocity (the root-mean-square of the turbulent kinetic energy) of the large scales convecting the small scales and the energy spectrum, but he also considered the effect of local straining of the small scale eddies on the correlation. Zhou and Rubenstein\cite{7} investigated both the non-local sweeping and local straining effects on the correlation to obtain the frequency spectra of sound using Lighthill's analogy. He et al.\cite{8} also showed that the sweeping velocity and the energy spectrum are essential ingredients for the use of large eddy simulation (LES) for the prediction of sound frequency spectra. However, this Eulerian formulation of Kraichnan resulted in a $k^{-3/2}$ wavenumber dependency for the kinetic energy spectrum in the inertial subrange, in disagreement with the Kolmogorov prediction\cite{9} of a $k^{-5/3}$ dependency. In a significant modification to his theory, which he called Lagrangian history direct interaction approximation (LHDIA), Kraichnan\cite{6,10} instead used Lagrangian space--time correlations of fluid particles as defined in Eq.~(\ref{eq2}) above. This modification resulted in agreement with the Kolmogorov $k^{-5/3}$ spectrum in the inertial subrange as well as with the associated Kolmogorov dissipation range universal spectrum. Furthermore, LHDIA agrees with Taylor's\cite{2} analysis of dispersion of a single particle and, for flows with an inertial subrange, agrees with Richardson's\cite{11} analysis describing the dispersion of two particles in a turbulent field. Further modifications to LHDIA were made by Kraichnan and Herring\cite{12} by considering Lagrangian correlations of the strain-rate field rather than the velocity field.

Lagrangian correlation functions play a role in the Lagrangian subgrid-scale LES model for turbulent flows by Meneveau et al.\cite{13} They used the dynamic procedure of obtaining Smagorinsky eddy-viscosity model coefficients for the subgrid-scale field from the resolved field but averaged the coefficients over Lagrangian pathlines, allowing the model to be readily used for inhomogeneous flows. The averaging times were determined from the Lagrangian correlation functions.

Bernard et al.\cite{14} and Bernard and Handler\cite{15} analyzed momentum transport in a turbulent channel flow and showed that the Reynolds shear stress could be decomposed into what they called displacement and acceleration transport terms, respectively, the former being of the mean gradient type and the latter being of the counter-mean gradient type. Furthermore, they showed that the eddy viscosity coefficient of the displacement transport term is properly expressed as a Lagrangian integral time scale obtained from the Lagrangian correlation of an ensemble of particles, as expressed by Eq.~(\ref{eq2}). Cho et al.\cite{16} used the tensorial Lagrangian time scales obtained from Eq.~(\ref{eq2}) in a new gradient transport model of the Reynolds stresses to represent the third order correlation functions. The time scales, determined from a channel flow DNS, were found to be different in the different coordinate directions.

In a very interesting paper, Phillips\cite{17} theoretically constructed a generic form of Eulerian space--time correlations of velocity component fluctuations applied to turbulent shear flows. The basis of his ideas originate with what is called the Kovasznay--Corrsin conjecture that, for homogeneous isotropic turbulence, space--time correlations can be expressed as spatial correlations and their dimunition with time. Among other results, Phillips\cite{17} defined a half-width of component correlation functions that collapse all the $R_{ii}$ component data from Kim and Hussain,\cite{18} described below, of the correlations for optimum time delay. Furthermore, he derived a generic expression for the convection velocities of the velocity component fluctuations that also compares well with DNS determined distributions of Kim and Hussain.\cite{18}

He and Zhang\cite{19} have formulated an elliptic model for Eulerian space--time correlations
for flows with mean shear, $\overline{U}(y)$, using a Taylor series expansion. Kaneda
and Gotoh\cite{20} and Kaneda\cite{21} previously used a Taylor series expansion in their analysis of both Eulerian and Lagrangian correlation functions in isotropic flow. The elliptic model of He and Zhang\cite{19} relates correlations, with spatial separations as the only independent variable, to space--time correlations by using two characteristic velocities, i.e., a convection velocity and a ``sweeping'' velocity that depends on the turbulence intensity and shear rate. This sweeping velocity is related to Kraichnan's\cite{6} ``sweeping hypothesis'' idea. They point out that Taylor's frozen turbulence hypothesis\cite{22} uses only one characteristic velocity, i.e., the convection velocity. When Taylor's hypothesis is invoked, the space--time correlation, for separation distances $r$ in the streamwise direction , can be expressed as $R(r, \tau) = R(r - U_{\rm c} \tau, 0)$, which assumes a linear space--time transformation and implies that the isocorrelation contours of this function are straight lines, $r - U_{\rm c} t = C$, where $C$ depends on the contour level, as shown in Fig.~\ref{fig1}. This had been previously been noted by Wills\cite{23} and clearly can not be true because correlations decay with increasing time and space separations. By contrast, the elliptic model describes the correlation at small separations as $R(r, \tau) = R(\sqrt{(r- U_{\rm c} {\tau})^2 + V^2 {\tau}^2}, 0)$, where the first term is the convection term and the sweeping velocity $V$ term comes from the Taylor expansion to second order. He and Zhang\cite{19} and Zhao and He\cite{24} have tested their model with a low Reynolds number turbulent channel flow (DNS) and found that the Eulerian space--time correlations collapse to a universal form throughout the flow, with the separation defined from the model, whereas this was only true in the outer part of the flow when Taylor's hypothesis was used. Furthermore, He et al.\cite{25} and Hogg and Ahlers\cite{26} have successfully applied this model in turbulent Rayleigh--Benard convection to convert temporal measurements into the spatial domain. He et al.\cite{27} extended this type of second order Taylor series analysis to Lagrangian space--time correlations in turbulent shear flows.

\begin{figure}[htbp]
\includegraphics{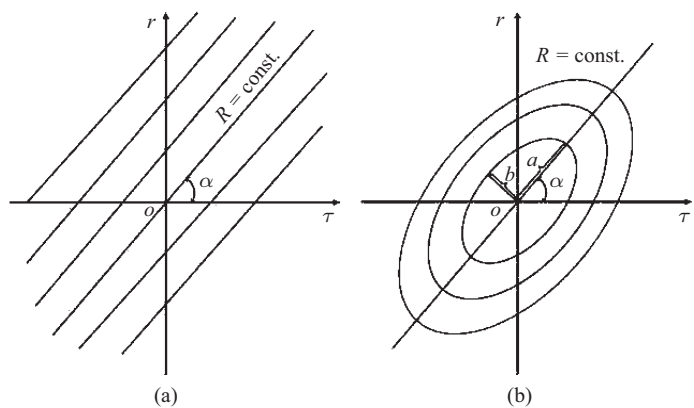}
\caption{\label{fig1} Space--time correlations of streamwise velocity fluctuations from turbulent channel flow with: (a) Taylor's frozen turbulence approximation\cite{22} and (b) the elliptic model of He and Zhang.\cite{19} Reprinted with permission of Zhao and He.\cite{24} (Copyright \copyright~2009 Am. Phys. Soc.)}
\end{figure}

\section{Eulerian experiments and simulations}

For most of the earlier period under review, with rare exception, experiments using space--time correlations were done in an Eulerian frame of reference. This was the case for the simple reason that the necessary particle tracking, required for a Lagrangian frame of reference, was nearly impossible experimentally with the technology available at the time. Numerical simulations were not available because of similar limitations of computer technology. For adequate particle tracking, experiments require advanced optical technology and computer imaging techniques, and numerical studies require high temporal and spatial resolution. These possibilities only became available and practical rather recently.

A series of the earliest and most influential of experimental investigations using Eulerian
space--time correlations of velocity fields were carried out by a research group at the Institut de M\'{e}canique Statistique de la Turbulence of the University of Marseille, France in the late 1940s and 1950s.\cite{28,29,30,31} The space--time correlation analysis was made using an analog recording of the signals on magnetic tape from hot-wire probes used to measure the streamwise velocity at two locations in the flow. These signals were played back with time shifts with respect to each other to vary $\tau$ in Eq.~(\ref{eq1}). Many of their results are summarized by Favre et al.\cite{32} and reviewed by Favre.\cite{33} Space--time correlations can be used to determine convection velocities for the turbulent fluctuations, as illustrated by Fig.~\ref{fig2} from their paper in 1962.\cite{32} Contours of constant correlation coefficient are shown where, for this figure, the indices and terms from Eq.~(\ref{eq1}) are $i = j = 1$, $x_2 = x_3 = \Delta x_2 = \Delta x_3 = 0$, and the axes are labeled with $ X_1 = \Delta x_1$, $T = \tau$, the grid cell size $M$, and the mean velocity $V$ for this grid turbulence study. The slope of the locus of maximum correlation along the diagonal ridge in Fig.~\ref{fig2} is the convection velocity of the streamwise velocity in the flow, as a function of downstream distance. This convection velocity is equal to the local mean velocity for grid flow. The authors note that the contours of constant correlation are elliptical.
\begin{figure}[htbp]
\includegraphics{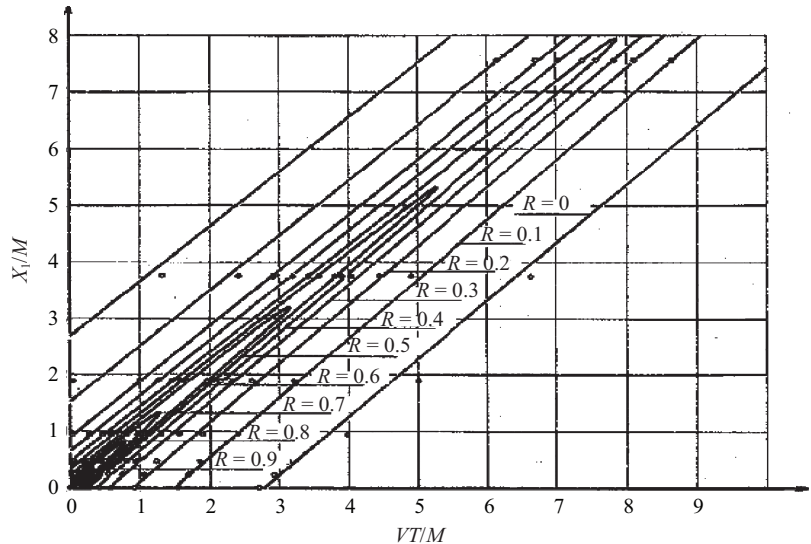}
\caption{\label{fig2} Eulerian space--time correlation of streamwise velocity fluctuations from wind-tunnel grid flow. Reprinted with permission of Favre et al.\cite{32} (Copyright \copyright~1962 CNRS)}
\end{figure}

This type of data can be displayed in a different format for discrete streamwise separation
distances. Figure~\ref{fig3} has the correlation coefficient as the vertical axis and $VT/M$ as the horizontal axis, and each curve corresponds to a discrete separation of the two probes, $\Delta x_1 = X_1$. The line through the maxima of the correlation coefficient curves corresponds to the ridge along the locus of the correlation maxima of Fig.~\ref{fig2}, and it illustrates the diminution of the correlation of the two signals with increasing separation between them. Additionally, these authors bandpass filtered the data in order to prescribe the convection velocity for a narrower range of turbulence scales. Results with bandpass filtering were reported in more detail by Favre et al.\cite{34}

\begin{figure}[htbp]
\includegraphics{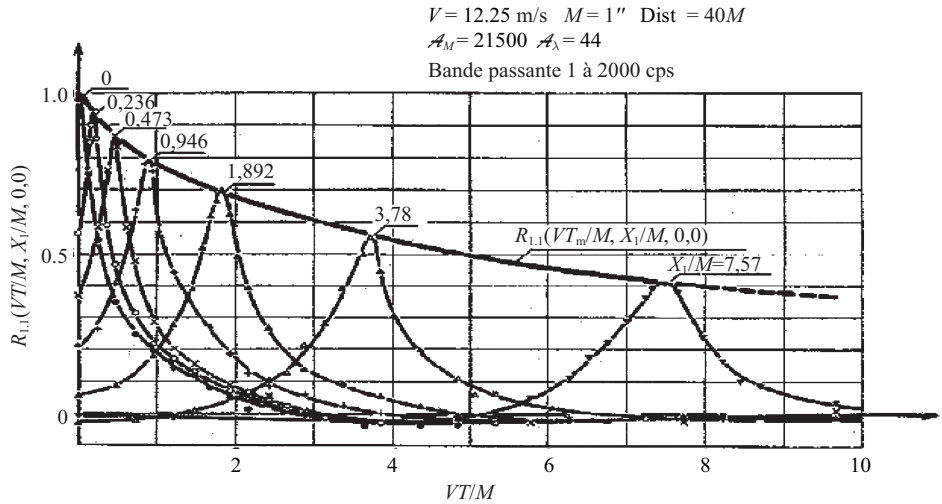}
\caption{\label{fig3} Eulerian correlation with time delay of streamwise velocity fluctuations from two locations separated by discrete distances in the streamwise direction in wind-tunnel grid flow. Reprinted with permission of Favre et al.\cite{32} (Copyright \copyright~1962 CNRS)}
\end{figure}

In a shear flow, when the variable measurement location is displaced throughout the flow
relative to the fixed measurement location, the shape and extent of the iso-correlation contours give some indication of the shape and size of the flow structures underlying the correlation. For example, Favre et al.\cite{35,36} did turbulent boundary layer experiments with hot-wire probes to measure streamwise velocity fluctuations, where one probe had stationary locations near the wall, and the other probe was moved to locations throughout the streamwise plane ($x$--$y$) and the cross-stream ($y$--$z$) planes, respectively, to determine space--time correlation coefficient contours. Figure~\ref{fig4} illustrates their results. Note that their coordinate system is labeled as $X_1 = x$, $X_3 = y$, and $X_2 = z$. In the lower part of the figure the iso-correlation curves in the streamwise plane are shown with the thick solid line drawn through the locus of maximum correlation. This line is inclined away from the wall in both the upstream and downstream directions indicating an average structure with such inclinations. Notably, too, the correlation levels remain relatively high for large distances away from the fixed probe. The iso-correlation contours are elongated in the streamwise direction, indicating structures of large extent in that direction. The four plots in the upper part of the figure are cross-stream sections at the streamwise locations indicated. In these cross-stream ($y$--$z$) planes, the correlation contours are longer in the wall-normal than in the spanwise direction, indicating relatively narrow average structures.

\begin{figure}[htbp]
\includegraphics{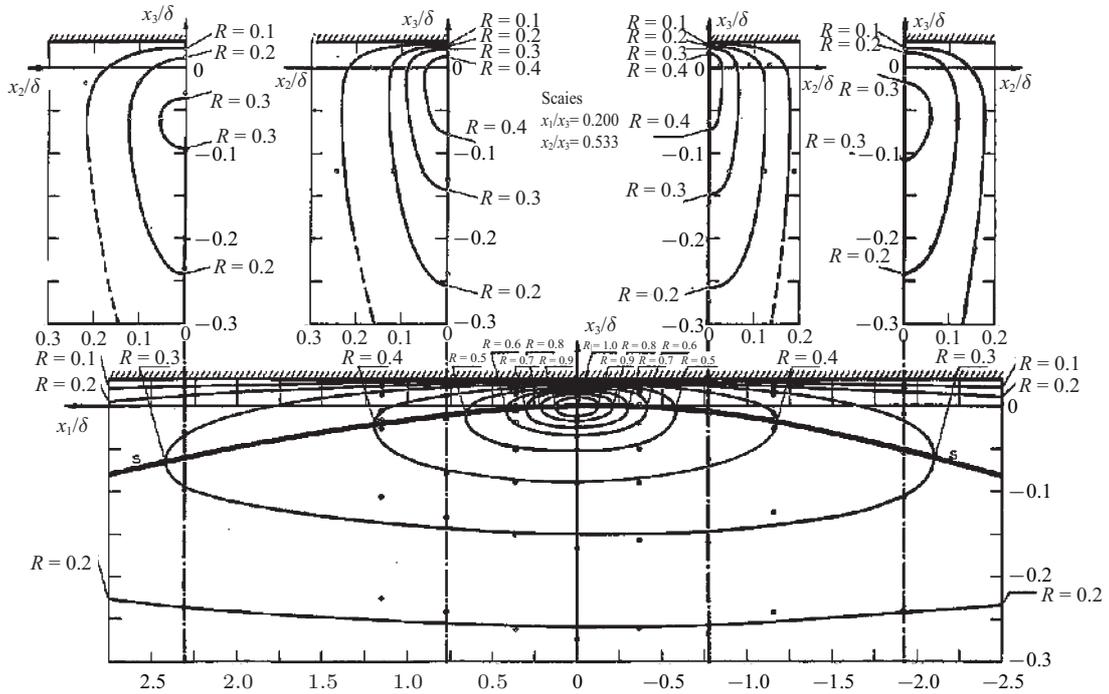}
\caption{\label{fig4} Space--time correlation of streamwise velocity fluctuations in a turbulent boundary layer, with the fixed probe at $y/\delta = 0.03$. Reprinted with permission of Favre et al.\cite{32} (Copyright \copyright 1962 CNRS)}
\end{figure}

Numerous other experimental and, later, direct numerical simulation studies have employed
space--time correlations, beginning in the 1960s. For example, Willmarth and Wooldridge\cite{37} made measurements at the wall of a turbulent boundary layer with a movable pressure sensor separated in the streamwise direction from a fixed upstream pressure sensor. Figure~\ref{fig5} shows their three-dimensional plot with the correlation coefficient as the vertical axis and the streamwise separation $\Delta x = x_1$ and the time delay $\tau$ as the horizontal axes. The dimunition of correlation with increasing probe separation and time delay is due to the increasing loss of contribution of small scales. The convection velocity was found to increase along the ridge of this plot, indicating that the larger scales propagate at a higher speed than the small scales. This larger propagation speed of the large scales seems plausible because their sources in the flow extend over greater wall normal distances and would be expected to travel, on average, with velocities of the flow further from the wall.
Willmarth and Yang\cite{38} compared these planar boundary layer results to new ones they obtained for the boundary layer over a cylindrical surface with the axis of symmetry in the streamwise direction.

\begin{figure}[htbp]
\includegraphics{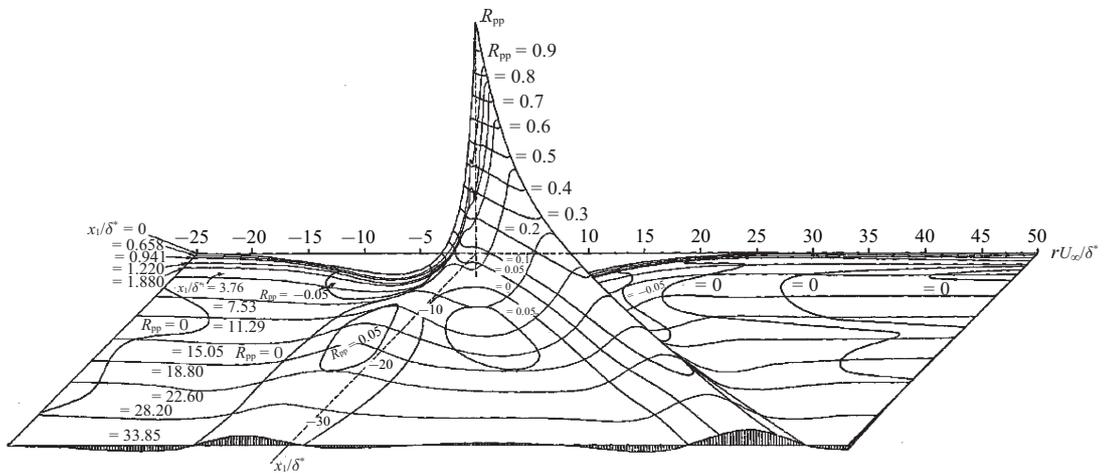}
\caption{\label{fig5} Space--time correlation of wall pressure fluctuations beneath a turbulent boundary layer. Reprinted with permission of Willmarth and Wooldridge.\cite{37} (Copyright \copyright~1962 Cambridge University Press)}
\end{figure}

Kistler and Chen\cite{39} made similar wall pressure measurements in a supersonic turbulent boundary layer with Mach numbers ranging between 1.33 and 5. Bull\cite{40} obtained broad and narrow frequency band wall pressure space--time correlations in a turbulent boundary layers with Mach numbers of 0.3 and 0.5 and a Reynolds number range of 5 to 1. He observed that the wall pressure appears to result from a variety of pressure sources in the flow with a wide range of convection velocities. They separate into two families, a high wavenumber, small scale group corresponding to the turbulence in the constant stress layer and a low wavenumber, large scale (twice the boundary layer thickness) group corresponding to the flow above this layer. The average convection velocity of the pressure fluctuations at the lower Mach number was 0.8 of the freestream velocity and fell to 0.6 at the higher Mach number.

Koplin\cite{41} made space--time correlation measurements of the streamwise velocity fluctuations in the mixing region of a subsonic turbulent jet, including bandpass filtering the data. He found that the shape of the correlations changed and the convection velocities decreased with increasing hot-wire sensor separation. He interpreted these changes to result from the fact that, at large separations, only information from the larger scales is contained in the correlation coefficients. Thus, in this unbounded flow, the relationship of the convection velocity to the scale of the turbulence appeared to be opposite to that for bounded flows. Fisher and Davies\cite{42} also made space--time correlation measurements in a subsonic turbulent jet, including bandpass filtering the data. They found that the convection velocity increased with an increase of the bandwidth center frequency, and also that the convection velocity was larger than the mean velocity in the outer half of the jet and less than the mean velocity in the inner half. Wills\cite{23} discussed space--time correlations in the context of turbulent jet flow experiments and analysis which extended Taylor's ``frozen turbulence'' hypothesis\cite{22} to account for the variable convection velocities for different scales of turbulence.

Eulerian space--time correlations were frequently used in turbulence investigations in the 1970s. Champagne et al.\cite{43} made such measurements of velocity and temperature fluctuations in homogeneous shear flow created in a wind tunnel with stacked flow conditioning channels of variable resistance. They also tested Taylor's\cite{22} hypothesis for this flow and found it to be sufficiently accurate. Comte-Bellot and Corrsin\cite{44} also employed narrow bandpass filtered space--time correlations to study the approximation of isotropic turbulence that is the flow downstream of a uniform grid. They were able to heuristically formulate a ``coherence time'' as a function of wavenumber that could be used to rescale the correlation delay times resulting in the filtered correlation coefficient curves collapsing into a single curve.

Kovasznay et al.\cite{45} and Blackwelder and Kovasznay\cite{46} made extensive space--time
correlation measuresments for all three fluctuating velocity components and the Reynolds shear stress in a turbulent boundary layer. Figure~\ref{fig6}(a), where the fixed probe was very close to the wall at $y/\delta = 0.03$ ($y^+ \approx 24$), shows the great extent above the wall and downstream within which the streamwise fluctuations are correlated. Here $y$ is the distance normal to the wall, $\delta$ is the boundary layer thickness and $y^+$ is $y$ normalized by the viscous length, $\nu/u_{\tau}$. Contrary to the results of Favre et al.,\cite{35,36} the isocorrelation contours are inclined only in the downstream direction with respect to the location of the fixed probe. From Fig.~\ref{fig6}(b), it is clear that the coherence of the wall normal fluctuations does not extend nearly so far above the wall and downstream as that of the streamwise fluctuations. This figure was obtained with the movable probe at the same \mbox{streamwise ($x$)} distance downstream as the fixed probe, but with varied wall-normal ($\Delta y$) and spanwise ($\Delta z$) spatial separations with respect to the fixed probe, as well as varied time delay ($\tau$). Sabot et al.\cite{47} extended such space--time correlation experiments to turbulent pipe flow, including measurements of radial velocity fluctuations.

\begin{figure}[htbp]
\includegraphics{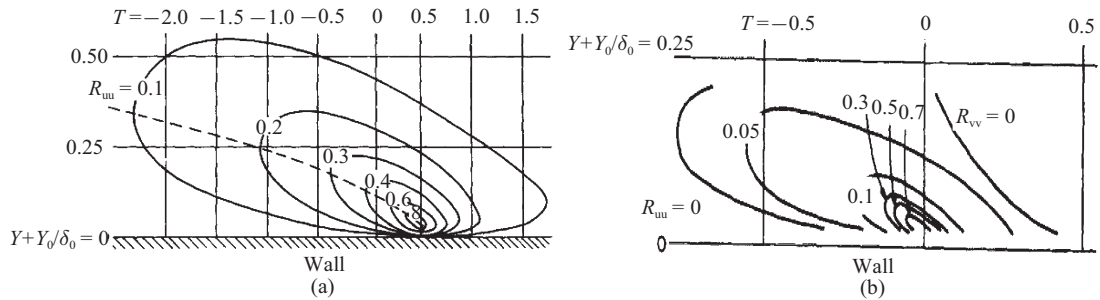}
\caption{\label{fig6} Space--time correlation of (a) streamwise and (b) wall-normal velocity fluctuations beneath a turbulent boundary layer. Reprinted with permission by Blackwelder and Kovasznay.\cite{46} (Copyright \copyright~1972 Cambridge University Press)}
\end{figure}

In the middle part of this decade Eulerian space--time correlation investigations were also extended by Demetriadesand\cite{48} to the compressible ($Ma=3.0$) axisymmetic wake of a circular cylindrical body to examine the flow structure and to a turbulent two-phase air-water mixture pipe flow with different inlet mixers by Herringe and Davis.\cite{49} At the end of the decade, Kreplin and Eckelmann\cite{50} made space--time correlation measurements of the streamwise and spanwise velocity fluctuations and their gradients at and normal to the wall. The experiment was carried out with hot-film sensors in a unique oil channel flow with a 5\;mm thick viscous sublayer, permitting measurements far deeper within the wall layer of a bounded flow than had ever been possible before. From these measurements they were able to construct an average picture of the flow structure near the wall made up of counter-rotating vortices inclined downstream at small angles to the wall that travel with a nearly constant convection velocity of about 12 times the friction velocity, $u_{\tau}$.

Although other experimental methods using conditional sampling and averaging were being developed in the 1970s, Eulerian space--time correlation methods continued to be used and extended in new ways. Goldschmidt et al.\cite{51} used broadband space--time correlations to show that convection velocities in a plane turbulent jet point outward from the streamlines and, from their bandpass results, that small scales convect at speeds greater than the local mean flow, while larger scales convect slower in agreement with Koplin's\cite{41} earlier results. Nagakawa and Nezu\cite{52} actually used the conditional averaging ideas that were developing in this decade to obtain conditionally averaged space--time correlations in an open channel flow. With one probe at a fixed location at the upper edge of the buffer layer and the other probe moved to variable locations with respect to the fixed probe position, they sorted the $u$ and $v$ product signals from the second and fourths quadrants of the Reynolds shear stress plane (see Ref.~\ocite{53}) to achieve this.

In the 1980s Smith and Townsend\cite{54} studied the structure of toroidal eddies in the Couette flow between two rotating concentric cylinders at high Taylor numbers. Among other things, they used an array of singe-sensor hot-wires, equally spaced in the direction parallel to the axes of the cylinders, to obtain space--time correlations. Bonnet et al.\cite{55} used two hot-wire probes to study the structure of the far wake developing downstream from turbulent boundary layers on both sides of the sharp trailing edge of a flat plate. They found that the double-roller structure observed in plane wakes originating from laminar boundary layers was not seen in their experiment. Bonnet\cite{56} revisited the study of wall pressure using space--time correlation, but with the added complexity of supersonic flow in a turbulent boundary layer with a shock-wave. Sirivat\cite{57} revisited the wind tunnel surrogate of isotropic flow, i.e., the flow downstream of a uniform grid. However, the novelty of this experiment was that the measurements were done with a single sensor hot-wire probe that moved with the flow by rotating it on a long arm. The validity of Taylor's hypothesis was confirmed and, importantly, a general expression for the correlation tensor with time delay was derived for isotropic flow, extending the K\'{a}rm\'{a}n-Howarth\cite{58} two-point correlation equation. Spina et al.\cite{59} used Eulerian space--time correlations to determine the average convection velocity of large scale structures in a supersonic boundary layer, obtaining a value of 0.9 times the freestream velocity throughout the outer part of the layer. They also found, by a pattern-recognition technique, that individual structures convect at approximately this velocity.

An important study was carried out by Kim and Hussain\cite{18} who used space--time correlations, obtained from a direct numerical simulation (DNS) of channel flow, to determine convection velocities of all three velocity and vorticity components, as well of wall pressure. They found that all these turbulence fluctuations convect at about the speed of the flow's local mean velocity for $y^+$ greater than about 20. Closer to the wall than this, they all converge to constant convection speeds of the order of about 10 times the friction velocity, in substantial agreement with Kreplin and Eckelmann.\cite{50} They also spatially filtered the data to study the dependence of the convection velocities on flow scale. They found little streamwise wavenumber ($k_x$) dependence; however, there is a strong spanwise wave number ($k_z$) dependence for $y^+ < 50$, with small scales convecting significantly slower than large scales. Romano\cite{60} performed an extensive study of Eulerian space, time and space--time correlations in a turbulent channel flow at several Reynolds numbers using laser-Doppler anemometry measurements at two locations separated in the streamwise direction. These measurements were highly resolved in space and time. Among other results, he confirmed the relationship between the convection velocity of the streamwise velocity fluctuations and local mean velocity previously found by Kim and Hussain.\cite{18} Romano\cite{60} found that high frequency fluctuations maintain their phase coherence more than their amplitude coherence as they convect downstream. Furthermore, so long as an optimized convection velocity is chosen, Romano concluded that the criteria for the applicability of Taylor's hypothesis can be expanded to $u^{\prime}/ \overline{U} < 0.3$ and $y^+ > 10$, where $u$ is the rms of the fluctuating streamwise velocity component and $\overline{U}$ is the local mean velocity component. Na and Moin\cite{61} examined the effects of mild and adverse pressure gradients on wall pressure for boundary layer direct numerical simulations. The adverse case resulted in separation with a closed separation bubble. From space--time correlations they found that the convection velocity of the pressure fluctuations decreases with increasing adverse pressure gradient and is quite reduced to a value as low as 55\%   the separation bubbleof the freestream velocity inside.

Using an electrochemical method to measure the streamwise velocity gradient fluctuations at the wall and laser doppler anemometry to measure the streamwise velocity fluctuations in the boundary layer flow above the wall, Labraga et al.,\cite{62} in an experiment very similar to that of Kreplin and Eckelmann,\cite{50} obtained space--time correlations of the two signals to determine angles of inclination with respect to the wall of the structures and their propagation velocities. Their results confirmed those of previous studies. Eulerian space--time correlations have even been used by Roy et al.\cite{63} to study the sizes and shapes of flow structures in a gravel-bed river field experiment using an array of electromagnetic current meters.

Motivated by the central role of space--time correlations in predicting sound generation from turbulent flows, and the attempts to make such predictions from LES calculations, He et al.\cite{8} investigated the effects of different subgrid scale models on the correlations in decaying isotropic turbulence. All of the models tested resulted in an under-prediction of the correlation magnitudes and a small over-prediction of the decorrelation time scales. Also motivated by the need to predict jet noise, Doty and McLaughlin\cite{64} made space--time correlation measurements of the radial density gradients in a jet shear layer at Mach numbers of 0.9 and 1.5 to demonstrate the strenghs and shortcomings of this unique measurement technique. Another supersonic flow study at Mach numbers of 2, 3, and 4 employing Eulerian space--time correlations was carried out by Barnardini and Pirozzoli.\cite{65} for DNS of turbulent boundary layers. They found that the correlations at supersonic Mach numbers closely resembled those in boundary layers at low speeds. Compressibility effects are quite weak. As for low-speed flows, the convection velocity of the low frequency pressure fluctuations was found to be about 80\% of the freestream velocity, whereas the convection velocities decrease systematically as the frequency increases.

\section{Lagrangian experiments and simulations}

Over a half century ago Durst et al.\cite{66} used Lagrangian correlations from geostrophic trajectories in horizontal planes to calculate the dispersion of fluid particles emitting from a point source. They found that the correlation coefficient followed an exponential distribution. Decades later P\'{e}cseli and Trulsen\cite{67} carried out a vortex method numerical study of geostropic flows from which they determined Eulerian and Lagrangian velocity correlations. In another early study without much experimental evidence to draw on, Philip\cite{68} developed and tested a relationship between Eulerian and Lagrangian correlation functions for isotropic turbulence with zero mean velocity.

Using the indirect method used by Townsend\cite{69} twenty years before, Schlien and Corrsin\cite{70} determined the Lagrangian correlation function from mean temperature profiles measured downstream of a heated wire, but with greater accuracy. These measurements were made in the same wind tunnel grid flow as the Eulerian correlation measurements of Comte-Bellot and Corrsin,\cite{44} so direct comparisons could be made for the same flow conditions. They found that the Lagrangian Taylor microscale was much larger than the corresponding Eulerian one. Almost twenty years later still, Karnik and Tavoularis\cite{71} extended such measurements to homogeneous shear flow. In these and other grid flow studies described below, the Eulerian space--time maximum correlation coefficient values determined from Eq.~(\ref{eq1}) at the values of ${\tau} = \Delta x/ \overline{U}$, illustrated by the correlation maxima envelop shown in Fig.~\ref{fig3}, can be meaningfully compared to the Lagrangian correlation coefficient values determined from Eq.~(\ref{eq2}) at variable values of $\tau$. Similar comparisons of Eulerian and Lagrangian space--time correlations can be made for other flows.

In one of the first experiments where it was attempted to measure Lagrangian correlations directly by photographing particle trajectories in the decaying wind tunnel turbulence downstream of a grid, Snyder and Lumley\cite{72} investigated the role of particle density by using small hollow glass particles, that were rather good surrogates of fluid particles, as well as a variety of heavier particles. They also measured the Eulerian streamwise velocity correlation using hot-wire anemometry. Within experimental accuracy, which was rather poor for the Lagrangian data, they found that the Eulerian time scale was roughly three times the Lagrangian one. In a breakthrough experiment over a decade and a half after the experiment of Snyder and Lumley,\cite{72} a time span that illustrates just how difficult such measurements were with the technology then available, Sato and Yamamoto\cite{73} used optical three-dimensional particle tracking (3D-PT) to determine the Lagrangian correlation function and the mean square lateral particle dispersion in the decaying and approximately isotropic turbulence of a water grid flow. They found that the Eulerian and Lagrangian correlation coefficient distributions were very similar when the time axis was scaled with the ratio of the Lagrangian to the Eulerian integral length scales. A much more recent particle tracking experiment is that of Guala and Liberzon\cite{74} in which they estimated Lagrangian correlations, using 3D-PT, of the rate of strain, enstrophy and their production rate terms in homogeneous turbulence at a Taylor scale Reynolds number of 50. Cross-correlation functions of these terms were also estimated. From the correlation functions, Lagrangian integral time scales were determined. In a very thorough new study of turbulent pipe flow, Oliveira et al.\cite{75} used 3D-PT with three cameras to track Lagrangian particle trajectories and their velocities and accelerations with high spatial and temporal resolution. They determined component Lagrangian velocity and acceleration auto- and cross-correlations at the highest shear flow Reynolds number to date, i.e., $Re_{\rm b} = 10\,300$ based on the bulk velocity and pipe diameter. They determined the Komogorov constant from analyses of their data and concluded that the small scales of this pipe flow are locally isotropic.

In one of the very earliest numerical attempts to determine single fluid particle and particle pair correlation functions and other Lagrangian statistics, and to compare Eulerian statistics with these as well as to obtain particle dispersion properties, Deardorff and Peskin\cite{76} analyzed data obtained from an LES of a turbulent channel flow. Among other results, they found that the two particle Lagrangian correlations were more persistent than those of a single particle. Recent studies\cite{77,78,79} found that an LES with the most widely used Smagorinsky SGS model could generate larger Lagrangian time scales than the ones in DNS. In another early numerical study of particle dispersion in turbulence, Riley and Patterson\cite{80} simulated isotropic flow in a 32$^3$ grid point calculation and tracked trajectories of fluid particle velocities, for the first time, by interpolation on the Eulerian grid. They also simulated the trajectories of rigid particles in this numerical flow. They found that, for the fluid particles and short times, the Lagrangian correlation decreased slower than the Eulerian correlation, but the opposite was true for large times. The Lagrangian correlation coefficient depended on the response time of the rigid particles to velocity changes. Squires and Eaton\cite{81} carried out DNS of decaying isotropic turbulence and of homogeneous shear flow. They determined the shapes of the Lagrangian correlation functions for all three velocity components of these flows and made comparisons to the Eulerian correlation coefficients. The value of the ratio of the Eulerian integral time scale to the Lagrangian integral time scale of $0.8$ that they found for the decaying isotropic case was in generally good agreement to this ratio found in the experiment of Sato and Yamamoto\cite{73} described above. Kuerten and Brouwers\cite{82} recently carried out a DNS of a turbulent channel flow at $Re_{\tau} = 950$, based on the friction velocity and channel half width, and determined Lagrangian statistics, including Lagrangian auto- and cross-correlation functions, which were compared to Langevin models.

With the first of numerous studies\cite{83,84,85,86,87} of Lagrangian flow properties by Yeung and co-investigators, Yeung and Pope\cite{83} extensively investigated statistics of velocity, acceleration, dissipation and other related properties in two DNS calculations (64$^3$ and 128$^3$ grid points) of isotropic turbulence with Taylor length scale Reynolds numbers of 38 to 93. Statistical stationarity was maintained by forcing the low wavenumber modes of the simulation. About 4\,000 particle trajectories were tracked providing Lagrangian velocity and velocity gradient values. Among the many results, they determined the Lagrangian correlation functions of acceleration and velocity magnitude over the range of Reynolds numbers, and, in addition, of the dissipation rate and enstropy. They also compared the Eulerian and Lagrangian velocity correlation functions and found that the former falls off slower than the latter with increasing time. The statistics of the dispersion of particle pairs in homogeneous shear flow were determined by Shen and Yeung,\cite{85} including two particle Lagrangian correlations as expressed by Eq.~(\ref{eq3}). Figure~\ref{fig7} illustrates such correlations with an initial particle separation with a magnitude of four Kolmogorov lengths but, for each curve, in one of the three coordinate directions. Yeung\cite{86} extended the isotropic turbulence study of Yeung and Pope\cite{83} to a Taylor scale Reynolds number of 234 with a 512$^3$ simulation. Yeung and Sawford\cite{87} examined the application of the ``random sweeping'' of small scales of turbulence, particularly for the scalar field, by the large scale motions. This type of study traces back to the ideas of Kraichnan,\cite{6} discussed above, and later of Tennekes.\cite{88}

\begin{figure}[htbp]
\includegraphics{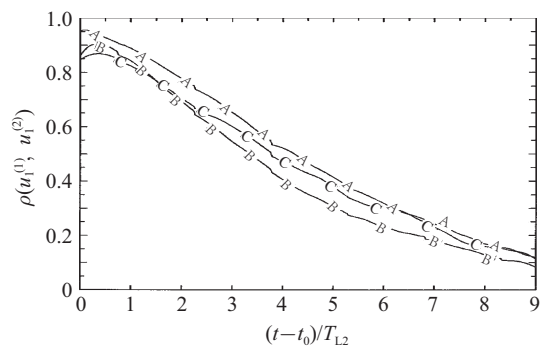}
\caption{\label{fig7} Two-particle Lagrangian correlations in homogeneous shear flow with an initial particle separation of four Kolmogorov lengths but, for each curve, in one of the three coordinate directions, $x$: A, $y$: B, and $z$: C. Reprinted with permission by Shen and Yeung.\cite{85} (Copyright \copyright~1977 Am. Inst. of Physics)}
\end{figure}

\section{Summary}

The information in turbulent fields at two points and two times, the separations of which can be varied, is rich. This information can be expressed statistically in space--time correlations, in both Eulerian and Lagrangian frames of reference, which play a central role in theories of turbulence and in attempts to model turbulence properties and processes. Such correlations also represent a type of experimental and numerical data analysis that has been, and continues to be, widely used in investigations of a variety of types of turbulent flows. With DNS investigations reaching ever higher Reynolds numbers and experimental investigations producing three-dimensional spatial data, both with the possibility of high spatial and temporal resolution, it is reasonable to assume that innovative new data analysis uses of space--time correlations and more complete tests of turbulence theories and models will be forthcoming.

%%%%%% DQLu

\vskip 5mm
{\small\textit{ %%
This review was supported by the Chinese Academy of Sciences and the Burgers Program
for Fluid Dynamics of the University of Maryland. It arose out of many fruitful discussions
with Professor Guowei He during my delightful visit, during the fall of 2013, to the State
Key Laboratory of Nonlinear Mechanics, Institute of Mechanics, Chinese Academy of Science,
which he directs. }}
} %\end {\setstretch{1.0}
%
%%\vskip 7mm %%%%%% DQLu

\end{document}

%% file: TAML-setting.tex
\usepackage{graphicx}% Include figure files  %%% \includegraphics
\usepackage{subfigure}  % ��ͼ Fig. 1a, 1b  %%% DQLu [2010.12.1]
\usepackage{dcolumn}% Align table columns on decimal point
\usepackage{bm}% bold math
\usepackage{fancyhdr}
\usepackage[pdfstartview=FitH,
            CJKbookmarks=true,
            bookmarksnumbered=true,
            bookmarksopen=true,
            colorlinks=true, %%ע�͵������򽻲�����Ϊ��ɫ�߿�(��colorlinks��pdfborderͬʱע�͵�)
            pdfborder=001,   %%ע�͵������򽻲�����Ϊ��ɫ�߿�
            linkcolor=blue,  %%�������Ϊ�ɫ
            citecolor=blue,  %%�������Ϊ�ɫ
            urlcolor=blue,   %%��ַ�͵���Ϊ�ɫ
            ]{hyperref}      %%�����

\renewcommand\footnoterule{\vspace{-0.1cm}\noindent\rule{5cm}{0.2pt}\vspace{1cm}} %% ��ע���߳�5cm����0.2pt

\renewcommand{\figurename}{Fig.}
\renewcommand{\tablename}{Table}

\renewcommand{\thetable}{\arabic{table}}

\def\ocite{\onlinecite} %% ����\onlinecite, ���ڴ��֣�������ֻ�ʺ� TAML ��
\newtoks\TitleOfPdf
\newtoks\ShortedTitle
\newtoks\ShortedAuthor
\newtoks\NumberOfPaper
\newtoks\VolNumberOfPaper
\newtoks\DoiOfPaper
\newtoks\Received
\newtoks\Revised
\newtoks\Accepted
\newtoks\PublishedOnline
\newtoks\ManuscriptProcessing   %%% ������������ʱ����
\newtoks\Logo
\newtoks\PaperHref
\PaperHref{\href{http://dx.doi.org/\the\DoiOfPaper}}
\if\mycmd1
\ManuscriptProcessing{{\small(Received \the\Received;  revised \the\Revised;  accepted \the\Accepted)}\\ \vskip -5pt}
\else
\ManuscriptProcessing{{\small(Received \the\Received;  accepted \the\Accepted)}\\ \vskip -5pt}
\fi

\Logo{\small \textcopyright\;{\footnotesize \textit{\the\year\, The Chinese Society of Theoretical and Applied
Mechanics. }}
  [doi:\the\PaperHref{\the\DoiOfPaper}]
   \\ \vskip -5pt}

\fancypagestyle{newpaper}{  %%%%%%% ���õ�һҳ��ҳü
\fancyhf{}
\fancyhead[C]{THEORETICAL \& APPLIED MECHANICS LETTERS \the\VolNumberOfPaper \hspace*{25mm}} %%%%%%% ÿƪ���µľ�����
}%

%% file: TAML-title.tex
%%%%%%%%%%%%%%%%%%%%%%%%%%%%%%%%%%%%%%%%%%%%%%%%%%%%%%%%%%%%%%%%%%%%%%%%%%%%%%%%%%%%%%%%%%%%%%%%%%%%%%%%%
%%%%%%%%%%%%%%%%%  Theoretical \& Applied Mechanics Letters                             %%%%%%%%%%%%%%%%%
%%%%%%%%%%%%%%%%%   2011 The Chinese Society of Theoretical and Applied Mechanics.      %%%%%%%%%%%%%%%%%
%%%%%%%%%%%%%%%%%  “一劳永逸”设置区   (题目及其链接、文档属性                         %%%%%%%%%%%%%%%%%
%%%%%%%%%%%%%%%%%%%%%%%%%%%%%%%%%%%%%%%%%%%%%%%%%%%%%%%%%%%%%%%%%%%%%%%%%%%%%%%%%%%%%%%%%%%%%%%%%%%%%%%%%

\hypersetup{ %%%%%%%%%%%%%%  PDF文件的属性 永久设置
            pdftitle={\the\TitleOfPdf}, %%%%%%%文章题目
            pdfauthor={\the\ShortedAuthor}, %%%%%%% 减缩的作者姓名
            pdfsubject={Theoretical \& Applied Mechanics Letters}  %
            }

\title{\the\PaperHref{ %%% Please do not input the Full title of paper in this line %%% 此处题目自动加链接
       \the\TitleOfPdf }   %%% 此处不要再输入题目
\\ \vskip -0.4cm ~ }

%%%%%%  DQLu [2011.5.10, Shanghai]  %%%%%%%%%%%%%%%%%%%%%%%%%%%%%%%%%%%%%%%%%%%%%%%%%%%%%%%%%%%%%%%%%%%%%%